\title{On the Limitations of the Anomalous Microwave Emission Emissivity}
\author{
        Christopher T. Tibbs\\
        \textit{Spitzer} Science Center\\
        Infrared Processing and Analysis Center\\
        California Institute of Technology\\
        Pasadena, CA 91125, \underline{USA}
            \and
        Roberta Paladini\\
        NASA \textit{Herschel} Science Center\\
         Infrared Processing and Analysis Center\\
        California Institute of Technology\\
        Pasadena, CA 91125, \underline{USA}      
            \and
        Clive Dickinson\\
        Jodrell Bank Centre for Astrophysics\\
        School of Physics and Astronomy\\
        The University of Manchester\\
        Manchester, M13 9PL, \underline{UK}      
}

\date{\today}
\documentclass[letterpaper,12pt, ]{article}

\usepackage[pdftex]{graphicx} 
\usepackage{natbib} 
\usepackage{multirow} 
\usepackage[caption=false]{subfig} 
\usepackage{enumerate} 
\usepackage[T1]{fontenc} 
\usepackage{amsmath} 

\def\reff@jnl#1{{\rm#1\/}}

\def\aj{\reff@jnl{AJ}}                  
\def\araa{\reff@jnl{ARA\&A}}            
\def\apj{\reff@jnl{ApJ}}                
\def\apjl{\reff@jnl{ApJ}}               
\def\apjs{\reff@jnl{ApJS}}              
\def\ao{\reff@jnl{Appl.Optics}}         
\def\apss{\reff@jnl{Ap\&SS}}            
\def\aap{\reff@jnl{A\&A}}               
\def\aapr{\reff@jnl{A\&A~Rev.}}         
\def\aaps{\reff@jnl{A\&AS}}             
\def\azh{\reff@jnl{AZh}}                        
\def\baas{\reff@jnl{BAAS}}              
\def\jrasc{\reff@jnl{JRASC}}            
\def\memras{\reff@jnl{MmRAS}}           
\def\mnras{\reff@jnl{MNRAS}}            
\def\pra{\reff@jnl{Phys.Rev.A}}         
\def\prb{\reff@jnl{Phys.Rev.B}}         
\def\prc{\reff@jnl{Phys.Rev.C}}         
\def\prd{\reff@jnl{Phys.Rev.D}}         
\def\prl{\reff@jnl{Phys.Rev.Lett}}      
\def\pasp{\reff@jnl{PASP}}              
\def\pasj{\reff@jnl{PASJ}}              
\def\qjras{\reff@jnl{QJRAS}}            
\def\skytel{\reff@jnl{S\&T}}            
\def\solphys{\reff@jnl{Solar~Phys.}}    
\def\sovast{\reff@jnl{Soviet~Ast.}}     
\def\ssr{\reff@jnl{Space~Sci.Rev.}}     
\def\zap{\reff@jnl{ZAp}}                        
\def\nat{\reff@jnl{Nature}}             

\begin{document}

\maketitle

\begin{abstract}

Many studies of anomalous microwave emission~(AME) have computed an AME emissivity to compare the strength of the AME detected in different regions. Such a value is usually defined as the ratio between the intensity of the AME at 1~cm and the thermal dust emission at 100~$\mu$m. However, as studies of Galactic dust emission have shown, the intensity of the thermal dust emission at 100~$\mu$m is strongly dependent on the dust temperature, which has severe implications for the AME emissivity defined in this way. In this work, we illustrate and quantify this effect and find that the AME emissivity decreases by a factor of 11.1 between dust temperatures of 20 and 30~K. We, therefore, conclude that computing the AME emissivity relative to the 100~$\mu$m emission does not allow for accurate comparisons between the AME observed in different environments. With this in mind, we investigate the use of other tracers of the dust emission with which to compute the AME emissivity and we ultimately conclude that, despite the difficulty in deriving its value, the column density of the dust would be the most suitable quantity with which to compute the AME emissivity.

\end{abstract}

\section{Introduction}

In recent years there has been growing evidence for the existence of a new component of microwave emission present in the interstellar medium~(ISM). This emission appears to be spatially correlated with the dust in the ISM, although it produces an excess of emission with respect to the predicted thermal, vibrational dust emission at these frequencies~\citep[e.g.][]{Leitch:97}, and as such this component is often described as anomalous. There have been only a handful of detections of this anomalous microwave emission (AME) component originating from both the diffuse ISM at mid-to-high latitudes~\citep[e.g.][]{Banday:03, Davies:06, Ghosh:12} and specific Galactic sources such as the Perseus and $\rho$ Oph molecular clouds~\citep{Watson:05, Casassus:08, Tibbs:10, Planck_Dickinson:11, Tibbs:12b}, the dark clouds LDN1622~\citep{Finkbeiner:02, Casassus:06}, LDN1621~\citep{Dickinson:10} and LDN1111~\citep{AMI_Scaife:09}, and a variety of HII regions~\citep{Dickinson:07, Todorovic:10, Tibbs:12a}. Additionally, the first detection of AME in a star forming region in the external galaxy, NGC6946, was reported~\citep{Murphy:10, AMI_Scaife:10}, confirming that this emission mechanism is truly ubiquitous.
 
Observations have shown that the AME occurs in the frequency range~$\sim$~10~--~100 GHz and is highly correlated with the mid-infrared~(IR) dust emission. It is this correlation with the mid-IR emission that led~\citet{DaL:98} to postulate their model of spinning dust emission. In this model, the observed excess emission is produced by the very smallest dust grains (Very Small Grains; VSGs or Polycyclic Aromatic Hydrocarbons; PAHs), characterised by an electric dipole moment, rotating rapidly, resulting in the production of electric dipole radiation. The model predicts a well defined peaked spectrum, rising below $\sim$~20~--~30~GHz and then falling off at higher frequencies as expected from a Boltzmann cutoff in grain rotation frequencies. The~\citet{DaL:98} spinning dust model has been refined and updated~\citep[][]{Ali-Haimoud:09, Hoang:10, Ysard:10, Hoang:11, Silsbee:11}, and current spinning dust models incorporate a variety of grain rotational excitation and damping processes: collisions with neutral and ionised gas particles, plasma drag (the interaction between the electric field of ions and the electric dipole moment of the dust grains), absorption and emission of a photon, the photoelectric effect, microwave emission and the formation of H$_{2}$ on the grain surface. Some of the more recent models incorporate additional features including dust grains that are not only spinning about their axis of greatest inertia~\citep{Hoang:10, Silsbee:11} and dust grains of irregular shape~\citep{Hoang:11}.

Understanding how the AME varies between different phases of the ISM is extremely important in improving our understanding of the AME. To help compare the strength of AME detected in one region to another, some authors have computed an AME emissivity. Given the strong association between the AME and the dust grains in the ISM, the AME emissivity is usually defined as the ratio between the antenna temperature of the AME at wavelengths of 1~cm and the surface brightness of the thermal dust emission at 100~$\mu$m. This quantity, first calculated as a cross-correlation coefficient over large areas of sky~\citep[e.g.][]{Kogut:96, deOC:97}, has now been computed for a variety of individual regions in which AME has been detected~(see Table~\ref{Table:Dust_Emissivity}). However, the AME emissivity defined in this manner is a highly biased method for making these comparisons because the surface brightness at 100~$\mu$m is significantly dependent on the dust temperature.

In this work we aim to (1) illustrate the weakness of such a definition and (2) present and discuss possible alternatives. In Section~\ref{sec:Dust_Emissivity} we point out why the commonly used AME emissivity definition is flawed and estimate the amplitude of the bias it introduces when comparing regions with different dust temperatures. In Section~\ref{sec:New} we discuss more suitable definitions of the AME emissivity including the use of gas and dust column densities. Finally, we present our conclusions in Section~\ref{sec:Conclusions}.

\begin{table}[!h]
\begin{center}
\caption{Sample of AME emissivity values for AME detections in a variety of different environments computed relative to the intensity of the 100~$\mu$m emission.}
\vspace{0.25cm}
\begin{tabular}{lcl}
\hline
Source & AME Emissivity & Reference \\
& ($\mu$K~(MJy~sr$^{-1}$)$^{-1}$) & \\
\hline
\hline

\textbf{H\textsc{ii} Regions} & & \\
6 southern H\textsc{ii} regions & 3.3~$\pm$~1.7  & \citet{Dickinson:07} \\  
9 northern H\textsc{ii} regions & 3.9~$\pm$~0.8 & \citet{Todorovic:10} \\ 
Pleiades & 2.01~$\pm$~0.09 & \citet{Genova-Santos:11} \\
RCW49 & 13.6~$\pm$~4.2 & \citet{Dickinson:07} \\
RCW175 & 14.2~$\pm$~2.7 & \citet{Tibbs:12a} \\
 & & \\

\textbf{High Latitudes} & & \\
15 regions \textit{WMAP} & 11.2~$\pm$~1.5 & \citet{Davies:06} \\
All-sky \textit{WMAP} & 10.9~$\pm$~1.1 & \citet{Davies:06} \\
 & & \\

\textbf{Molecular Clouds}  & & \\
Perseus & 15.7~$\pm$~0.3  & \citet{Watson:05} \\ 
Perseus A1 & 2.8~$\pm$~0.7 & \citet{Tibbs:10} \\ 
Perseus A2 & 16.4~$\pm$~4.1 & \citet{Tibbs:10} \\ 
Perseus A3 & 12.8~$\pm$~6.1 & \citet{Tibbs:10} \\ 
Perseus B & 13.2~$\pm$~3.6 & \citet{Tibbs:10} \\ 
Perseus C & 13.0~$\pm$~3.2 & \citet{Tibbs:10} \\ 
 & & \\

\textbf{Dark Clouds} & & \\
LDN1621 & 18.1~$\pm$~4.4 & \citet{Dickinson:10} \\
LDN1622 & 21.3~$\pm$~0.6 & \citet{Casassus:06} \\

\hline
\label{Table:Dust_Emissivity}
\end{tabular}
\end{center}
\end{table}

\section{The AME Emissivity Defined Relative to the 100~$\mu$m Emission}
\label{sec:Dust_Emissivity}

The concept of an AME emissivity is particularly useful as it provides a normalisation of the strength of the AME detected in a variety of regions, and hence it facilitates a comparison between such AME detections. In Table~\ref{Table:Dust_Emissivity} we list AME emissivities defined relative to the 100~$\mu$m emission from the literature for a variety of different phases of the ISM, including H\textsc{ii} regions, molecular clouds, dark clouds and diffuse emission at intermediate latitudes. Looking at the values in Table~\ref{Table:Dust_Emissivity} it is possible to identify the observed range of values of the AME emissivity. A typical value for the AME emissivity is of the order 10~$\mu$K~(MJy~sr$^{-1}$)$^{-1}$ with a range from $\sim$~3 to 25~$\mu$K~(MJy~sr$^{-1}$)$^{-1}$~\citep{Banday:03, Davies:06}. In terms of flux density, this corresponds to approximately 1~Jy of AME at 1~cm for every 3000~Jy at 100~$\mu$m. 

\begin{figure}
\begin{center}
\includegraphics[angle=0,scale=0.60]{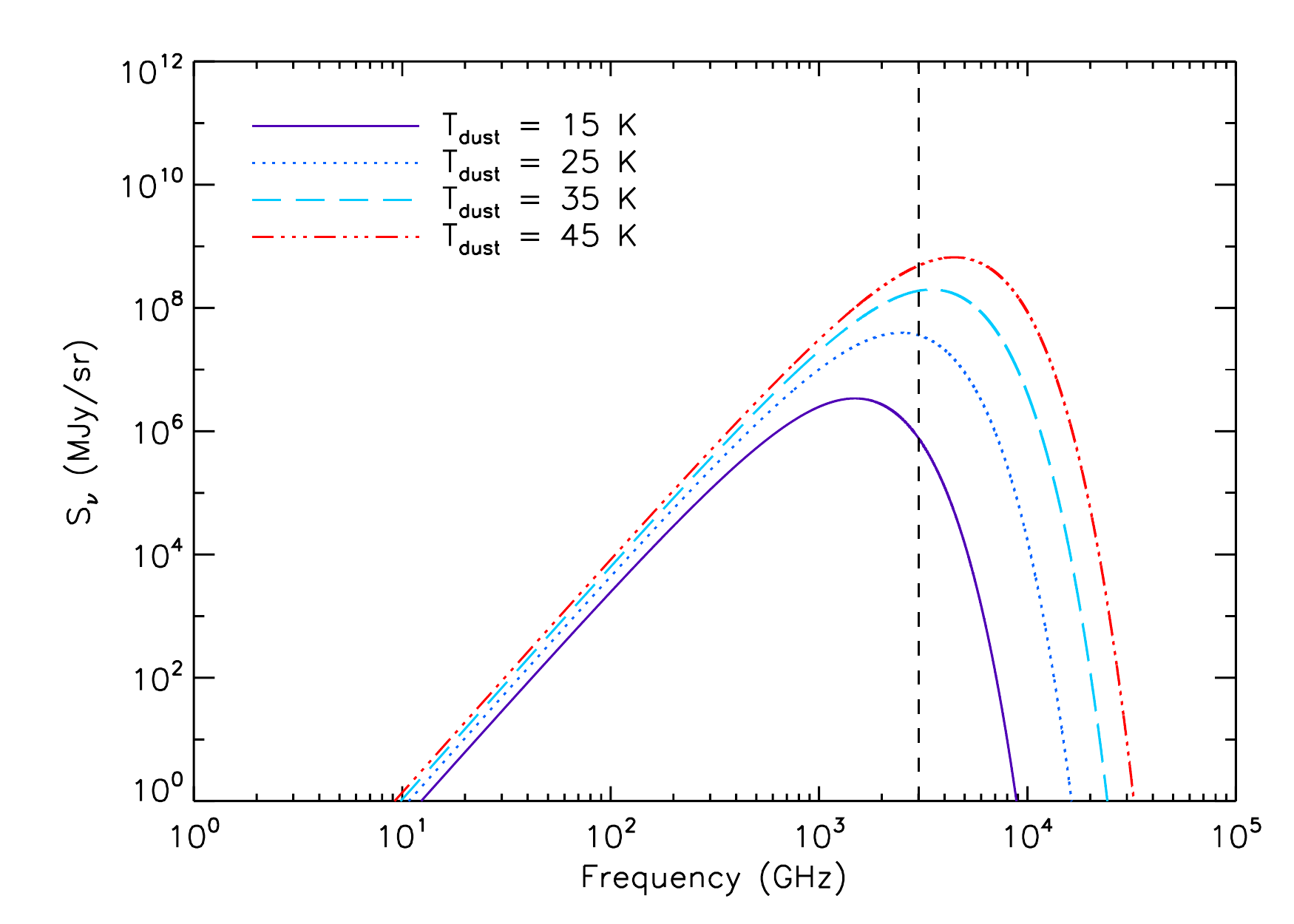}
\caption{The modified black body spectrum representing the thermal dust emission for a range of dust temperatures from 15 to 45~K with a fixed dust emissivity spectral index of 1.8. This illustrates how the 100~$\mu$m emission~(vertical dashed line) increases with increasing dust temperature, and hence how the AME emissivity defined relative to the 100~$\mu$m emission is biased by changes in dust temperature.}
\label{Fig:Emissivity_Td}
\end{center}
\end{figure}

Observations at far-IR and submillimeter wavelengths~\citep[e.g.][]{Reach:95, Boulanger:96} have revealed that the spectrum of the thermal dust emission is well represented by a modified black body function of the form 

\begin{equation}
S_{\nu} \propto \frac{2h\nu^{3+\beta}}{c^{2}} \frac{1}{\mathrm{exp}(h\nu/kT_{\mathrm{dust}}) -1}
\label{equ:bb}
\end{equation}

\noindent
where $\beta$ is the dust emissivity spectral index and T$_{\mathrm{dust}}$ is the equilibrium temperature of the big dust grains. The thermal dust emission observed at such far-IR and submillimeter wavelengths is produced by big dust grains absorbing ultra-violet photons from the exciting radiation field of the environment and re-radiating this emission thermally. It is this balance between absorption and emission that produces the well-defined spectrum of thermal dust emission~(see Figure~\ref{Fig:Emissivity_Td}) and keeps the big grains in thermal equilibrium. In the ISM, the dust temperature is known to vary from one environment to an other~\citep{Planck_DarkGas:11} with dust temperatures ranging from $\sim$~15~K in dense molecular clouds to 25~--~30~K in H\textsc{ii} regions. Additionally, the dust temperature has been observed to vary on small angular scales and within regions in which AME has been detected~\citep[e.g. between 15 and 25~K in the Perseus molecular cloud;][]{Tibbs:11}. Since $\beta$ is dependent on the physical properties of the dust grains, it is also expected to vary between different environments:~\citet{Dupac:03} found $\beta$ to vary from 0.8 to 2.4 in a range of Galactic environments with dust temperatures between 11 and 80~K. Therefore, given that the emission at 100~$\mu$m is an observed quantity that depends strongly T$_{\mathrm{dust}}$ and $\beta$, and since both T$_{\mathrm{dust}}$ and $\beta$ are known to vary in the different phases of the ISM, the 100~$\mu$m emission is not a suitable quantity with which to estimate the AME emissivity. 

To quantify the effect of dust temperature variations on the AME emissivity, we consider a dust emissivity spectral index of 1.8, which is the value derived for the Galactic solar neighbourhood~\citep{Planck_DarkGas:11}, and two dust temperatures~(20 and 30~K). We then calculate that for a given intensity at 1~cm, the AME emissivity at 20~K is 11.1 times higher than at 30~K. This effect is substantial and could by itself, without any intrinsic variation of the AME intensity, explain the variations of the AME emissivities listed in Table~\ref{Table:Dust_Emissivity}. This clearly demonstrates the bias in the AME emissivity when computed relative to the 100~$\mu$m emission. With this in mind, we can now revisit the AME emissivity values listed in Table~\ref{Table:Dust_Emissivity}. For example, we focus on the five regions of AME detected in the Perseus molecular cloud~(A1, A2, A3, B and C) by~\citet{Tibbs:10}. It is apparent that region A1 is much less emissive~(by a factor of~$\sim$~4~--~5) than the other four regions. However, as discussed by~\citet{Tibbs:10, Tibbs:11}, the physical environment of region A1, which corresponds to the open cluster IC348, is very different from the other four regions. In fact, simply because of the difference in dust temperature between IC348~($\sim$~22~K) and the other regions~($\sim$~18~K), we estimate that the AME emissivity is reduced by a factor of 4.3 in region A1 relative to the other regions. Therefore, the inconsistency between region A1 and the other regions can be solely accounted for by dust temperature variations, illustrating the impact of the dust temperature in biasing the AME emissivity computed using the 100~$\mu$m emission.

\section{Redefining the AME Emissivity}
\label{sec:New}

In Section~\ref{sec:Dust_Emissivity} we have shown that the commonly used definition of the AME emissivity computed relative to the 100~$\mu$m emission is flawed. Therefore, an unbiased definition of the AME emissivity is required to help quantify the AME variations.

As part of their investigation of the diffuse emission at intermediate latitudes,~\citet{Davies:06} computed the AME emissivity relative to the dust emission at 94~GHz rather than 100~$\mu$m. Since 94~GHz is further from the peak of the thermal dust emission (see Figure~\ref{Fig:Emissivity_Td}) the emission at 94~GHz is much less dependent on the dust temperature than the emission at 100~$\mu$m. However, even using the intensity of the 94~GHz emission, there is still an effect due to the dust temperature, as the AME emissivity decreases by a factor of 1.6 when the dust temperature increases from 20 to 30~K. 

Rather than using the intensity of the dust emission to compute an AME emissivity, other authors~\citep[e.g.][]{Finkbeiner:02, Vidal:11} have computed an AME emissivity relative to the hydrogen column density, N$_{\mathrm{H}}$. This definition has several advantages. First, it results in an AME emissivity that is more directly suited to comparing with the theoretical values as the current spinning dust theories compute an emissivity in units of Jy~sr$^{-1}$~cm$^{2}$ per H atom. Using N$_{\mathrm{H}}$ also mitigates the effects of dust temperature, hence allowing a much more impartial comparison between measurements of AME in different environments. To illustrate this, we use the N$_{\mathrm{H}}$ map of the Perseus molecular cloud from~\citet{Tibbs:11}, which is based on a visual extinction map, and compute the AME emissivity for the five regions of AME in the Perseus molecular cloud~(see Table~\ref{Table:Dust_Emissivity_NH}). When computing the AME emissivity relative to N$_{\mathrm{H}}$, we find that region A1 is now consistent~(within 2~--~3$\sigma$) with the four other regions in the Perseus molecular cloud, which leads to a completely different interpretation. We now want to explore how this new definition can help us to perform a less biased comparison between AME detections in different environments. \citet{Vidal:11} compared the AME emissivity of six regions covering a range of column densities from Galactic cirrus to dense dark clouds~(see Table~\ref{Table:Dust_Emissivity_NH}). Adding these six measurements to the five we computed for the Perseus molecular cloud, gives a sample of 11 AME emissivities based on N$_{\mathrm{H}}$. It is apparent that the AME emissivities for the Perseus molecular cloud are consistent with the values computed for other AME detections. It is also noticeable that there is less scatter in the AME emissivities listed in Table~\ref{Table:Dust_Emissivity_NH} compared to those listed in Table~\ref{Table:Dust_Emissivity}.

\begin{table}[!h]
\begin{center}
\caption{AME emissivity values for a variety of phases of the ISM computed relative to the hydrogen column density.}
\vspace{0.25cm}
\begin{tabular}{lccl}
\hline
Source & N$_{\mathrm{H}}$ & AME Emissivity & Reference \\
& (10$^{22}$ H~cm$^{-2}$) & (10$^{-18}$ Jy~sr$^{-1}$~cm$^{2}$/H) & \\
\hline
\hline

Perseus A1 & 0.82~$\pm$~0.19 & 2.2~$\pm$~0.5 &This work \\
Perseus A2 & 1.19~$\pm$~0.27 & 2.7~$\pm$~0.6 & This work \\
Perseus A3 & 1.28~$\pm$~0.29 & 1.2~$\pm$~0.3 & This work \\
Perseus B & 0.79~$\pm$~0.18 & 3.4~$\pm$~0.8 & This work \\
Perseus C & 0.74~$\pm$~0.17 & 3.6~$\pm$~0.8 & This work\\
\\
Cirrus & 0.15~$\pm$~0.07 & 4.6~$\pm$~2.0 & \citet{Leitch:97} \\
$\zeta$ Oph & 0.22~$\pm$~0.02 & 4.1~$\pm$~0.6 & \citet{Vidal:11} \\
LDN1780 & 0.45~$\pm$~0.04 & 3.5~$\pm$~0.4 & \citet{Vidal:11} \\
LDN1622 & 1.50~$\pm$~0.15 & 2.0~$\pm$~0.2 & \citet{Casassus:06} \\
$\rho$ Oph & 5.00~$\pm$~0.50 & 3.2~$\pm$~0.5 & \citet{Casassus:08} \\
M78 & 22.80~$\pm$~0.23 & 0.9$~\pm$~0.1 & \citet{Castellanos:11} \\

\hline
\label{Table:Dust_Emissivity_NH}
\end{tabular}
\end{center}
\end{table}

\begin{figure}
\begin{center}
\includegraphics[angle=0,scale=0.60]{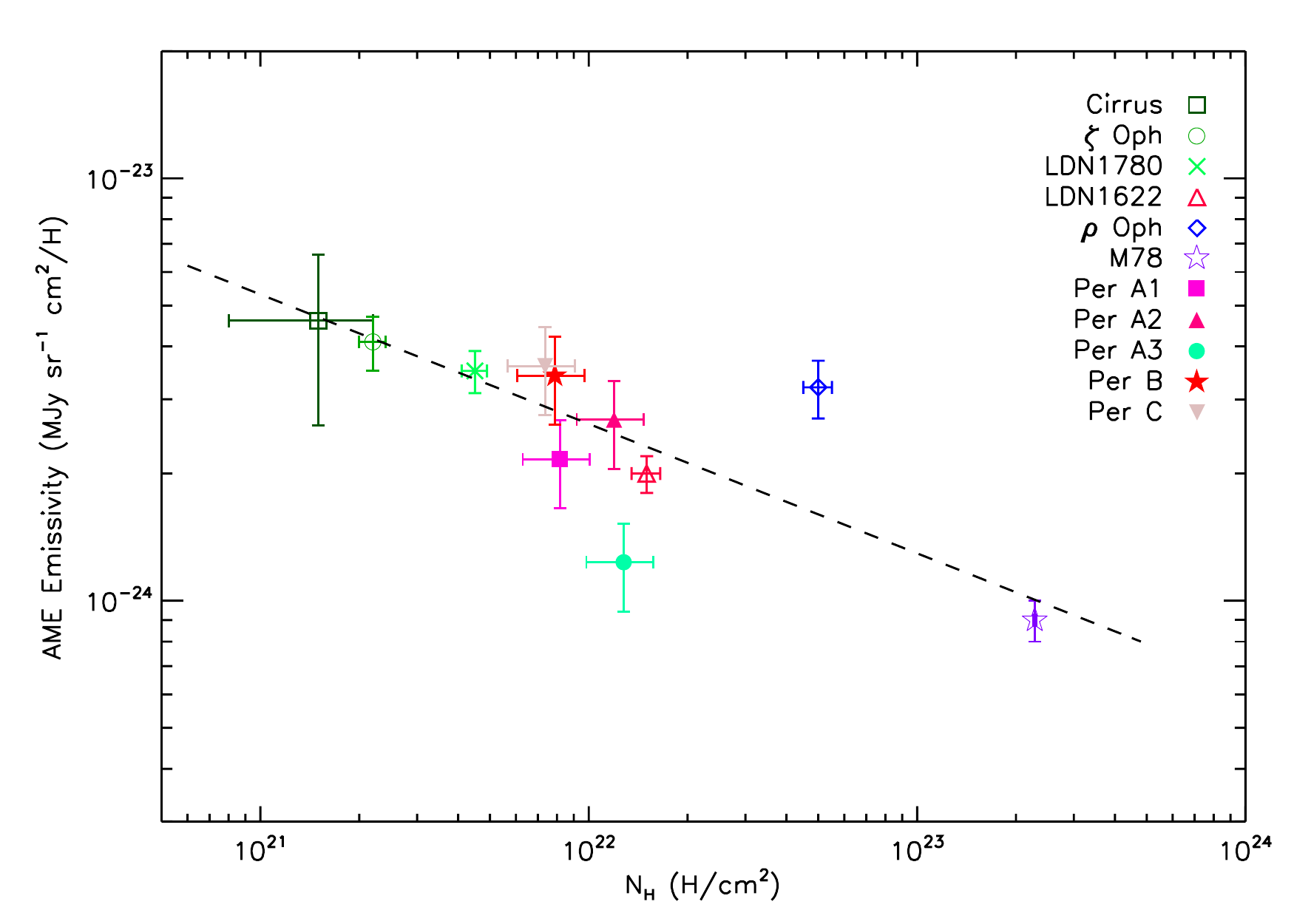} 
\caption{AME emissivity computed using N$_{\mathrm{H}}$ for the AME detections listed in Table~\ref{Table:Dust_Emissivity_NH} plotted as a function of N$_{\mathrm{H}}$. Also plotted is the fit to the data~(dashed line), which we estimate to have a spectral index of $-$0.31~$\pm$~0.03.}
\label{Fig:Emissivity_NH}
\end{center}
\end{figure}

Following the example of~\citet{Vidal:11}, we plotted the AME emissivities listed in Table~\ref{Table:Dust_Emissivity_NH} as a function of N$_{\mathrm{H}}$. This plot is displayed in Figure~\ref{Fig:Emissivity_NH}. We fitted the data with a power-law and found that the best fitting spectral index is $-$0.31~$\pm$~0.03. Although there is some scatter in the data, the results hint that the AME emissivity is decreasing with increasing N$_{\mathrm{H}}$, suggesting that the AME becomes less emissive as the column density increases. This is consistent with the AME being due to the smallest dust grains, as it is known that as N$_{\mathrm{H}}$ increases, the abundance of the smallest dust grains decreases due to dust grain coagulation~\citep[e.g.][]{Stepnik:03, Flagey:09}. The idea of the AME being associated with the smallest dust grains is additional support for the AME being due to spinning dust.

Although we have shown that defining the AME emissivity using N$_{\mathrm{H}}$ removes the bias introduced by the dust temperature and allows for a much more impartial comparison of AME between different environments, estimating N$_{\mathrm{H}}$, including molecular, atomic and ionised forms, is technically challenging. There are three methods used to estimate N$_{\mathrm{H}}$: near-IR extinction mapping; far-IR thermal dust emission; and the use of gas tracers. However, each method has its limitations. For example, near-IR extinction is only reliable if there are a sufficient number of background sources available, thermal dust emission estimates are affected by variations in the dust opacity, while observations of gas tracers, although they do provide kinematical information, are limited by the critical density of the observed molecule. A detailed analysis performed by~\citet{Goodman:09} clearly illustrates the various uncertainties and limitations in estimating N$_{\mathrm{H}}$ using the three separate methods. 

In addition to the difficulties in measuring N$_{\mathrm{H}}$, the AME is believed to be associated with interstellar dust grains, and hence it would make more sense to normalise the strength of the AME with a physical property of the dust. For example, the column density of dust, N$_{\mathrm{dust}}$, would be a useful quantity as it has the same advantages as using N$_{\mathrm{H}}$ with the addition that it is a direct tracer of the dust. However, N$_{\mathrm{dust}}$ is even more difficult to estimate than N$_{\mathrm{H}}$. It is possible to assume a given M$_{\mathrm{dust}}$/M$_{\mathrm{H}}$ and convert from N$_{\mathrm{H}}$ to N$_{\mathrm{dust}}$, however this assumes that we know M$_{\mathrm{dust}}$/M$_{\mathrm{H}}$. Globally, within our own Galaxy, it is known that M$_{\mathrm{dust}}$/M$_{\mathrm{H}}$ is~$\sim$~0.01~\citep[e.g.][]{DraineLi:07}, but AME has recently been observed in external galaxies~\citep[e.g.][]{Murphy:10, Planck_SMC:11} where it is known that the metallicity, and hence M$_{\mathrm{dust}}$/M$_{\mathrm{H}}$, differs from our own Galaxy~\citep[e.g.][]{Draine:07}. To mitigate the affects introduced by variations in M$_{\mathrm{dust}}$/M$_{\mathrm{H}}$, N$_{\mathrm{dust}}$ may be computed by fitting the far-IR SED with a dust model which incorporates the physical properties of the dust grains. However, to date, no such modelling has yet been performed systematically for the environments in which AME has been detected.

\section{Conclusions}
\label{sec:Conclusions}

Having discussed various definitions of the AME emissivity, it is clear that using the 100~$\mu$m emission introduces a bias due to the effect of dust temperature, and that using N$_{\mathrm{H}}$ allows for more accurate comparisons. However, given the association between the AME and the interstellar dust, an intrinsic property of the dust such as N$_{\mathrm{dust}}$, would represent the best quantity with which to compute the AME emissivity. Therefore, until such a time as a new AME emissivity is defined, we stress that care should be taken before using it to directly compare the strength of AME in different environments.

\vspace{1.5cm}
\noindent
We thank the referee for useful comments which helped improve the content of the paper. This work has been performed within the framework of a NASA/ADP ROSES-2009 grant, no. 09-ADP09-0059. CD acknowledges support from an SFTC Advanced Fellowship and an EU Marie-Curie IRG grant under the FP7.



\begin{thebibliography}{}

\bibitem[Ali-Ha{\"i}moud, Hirata \& Dickinson(2009)]{Ali-Haimoud:09} 
Ali-Ha{\"i}moud Y., Hirata C.~M., Dickinson C.,\ 2009, \mnras, 395, 1055 

\bibitem[AMI Consortium: Scaife et al.(2009)]{AMI_Scaife:09} 
AMI Consortium: Scaife, A.~M.~M., Hurley-Walker, N., et al.\ 2009, \mnras, 394, L46 

\bibitem[AMI Consortium: Scaife et al.(2010)]{AMI_Scaife:10} 
AMI Consortium: Scaife, A.~M.~M., Nikolic, B., Green, D.~A., et al.\ 2010, \mnras, 406, L45 

\bibitem[Banday et al.(2003)]{Banday:03} 
Banday, A.~J., Dickinson, C., Davies, R.~D., Davis, R.~J., \& G{\'o}rski, K.~M.\ 2003, \mnras, 345, 897 

\bibitem[Boulanger et al.(1996)]{Boulanger:96} 
Boulanger, F., Abergel, A., Bernard, J.-P., et al.\ 1996, \aap, 312, 256 

\bibitem[Casassus et al.(2006)]{Casassus:06} 
Casassus, S., Cabrera, G.~F., F{\"o}rster, F., et al.\ 2006, \apj, 639, 951 

\bibitem[Casassus et al.(2008)]{Casassus:08} 
Casassus, S., Dickinson, C., Cleary, K., et al.\ 2008, \mnras, 391, 1075

\bibitem[Castellanos et al.(2011)]{Castellanos:11} 
Castellanos, P., Casassus, S., Dickinson, C., et al.\ 2011, \mnras, 411, 1137 

\bibitem[Davies et al.(2006)]{Davies:06} 
Davies, R.~D., Dickinson, C., Banday, A.~J., et al.\ 2006, \mnras, 370, 1125 

\bibitem[de Oliveira-Costa et al.(1997)]{deOC:97} 
de Oliveira-Costa, A., Kogut, A., Devlin, M.~J., et al.\ 1997, \apjl, 482, L17 

\bibitem[Dickinson et al.(2007)]{Dickinson:07} 
Dickinson, C., Davies, R.~D., Bronfman, L., et al.\ 2007, \mnras, 379, 297 

\bibitem[Dickinson et al.(2010)]{Dickinson:10} 
Dickinson C. et al.,\ 2010, \mnras, 407, 2223 

\bibitem[Draine \& Lazarian(1998)]{DaL:98} 
Draine B.~T., Lazarian A.,\ 1998, \apj, 508, 157 

\bibitem[Draine \& Li(2007)]{DraineLi:07} 
Draine, B.~T., \& Li, A.\ 2007, \apj, 657, 810 

\bibitem[Draine et al.(2007)]{Draine:07} 
Draine, B.~T., Dale, D.~A., Bendo, G., et al.\ 2007, \apj, 663, 866 

\bibitem[Dupac et al.(2003)]{Dupac:03} 
Dupac, X., Bernard, J.-P., Boudet, N., et al.\ 2003, \aap, 404, L11 

\bibitem[Finkbeiner et al.(2002)]{Finkbeiner:02} 
Finkbeiner, D.~P., Schlegel, D.~J., Frank, C., \& Heiles, C.\ 2002, \apj, 566, 898

\bibitem[Flagey et al.(2009)]{Flagey:09} 
Flagey, N., Noriega-Crespo, A., Boulanger, F., et al.\ 2009, \apj, 701, 1450 

\bibitem[G{\'e}nova-Santos et al.(2011)]{Genova-Santos:11} 
G{\'e}nova-Santos, R., Rebolo, R., Rubi{\~n}o-Mart{\'{\i}}n, J.~A., L{\'o}pez-Caraballo, C.~H., \& Hildebrandt, S.~R.\ 2011, \apj, 743, 67 

\bibitem[Ghosh et al.(2012)]{Ghosh:12} 
Ghosh, T., Banday, A.~J., Jaffe, T., et al.\ 2012, \mnras, 422, 3617 

\bibitem[Goodman et al.(2009)]{Goodman:09} 
Goodman, A.~A., Pineda, J.~E., \& Schnee, S.~L.\ 2009, \apj, 692, 91 

\bibitem[Hoang et al.(2010)]{Hoang:10} 
Hoang, T., Draine, B.~T., \& Lazarian, A.\ 2010, \apj, 715, 1462

\bibitem[Hoang et al.(2011)]{Hoang:11} 
Hoang, T., Lazarian, A., \& Draine, B.~T.\ 2011, \apj, 741, 87 

\bibitem[Kogut et al.(1996)]{Kogut:96} 
Kogut, A., Banday, A.~J., Bennett, C.~L., et al.\ 1996, \apjl, 464, L5 

\bibitem[Leitch et al.(1997)]{Leitch:97} 
Leitch, E.~M., Readhead, A.~C.~S., Pearson, T.~J., \& Myers, S.~T.\ 1997, \apjl, 486, L23 

\bibitem[Murphy et al.(2010)]{Murphy:10} 
Murphy, E.~J., Helou, G., Condon, J.~J., et al.\ 2010, \apjl, 709, L108 

\bibitem[Planck Collaboration(2011a)]{Planck_DarkGas:11} 
Planck Collaboration,\ 2011a, \aap, 536, A19

\bibitem[Planck Collaboration(2011b)]{Planck_Dickinson:11} 
Planck Collaboration,\ 2011b, \aap, 536, A20 

\bibitem[Planck Collaboration(2011c)]{Planck_SMC:11} 
Planck Collaboration,\ 2011c, \aap, 536, A17 

\bibitem[Reach et al.(1995)]{Reach:95} 
Reach, W.~T., Dwek, E., Fixsen, D.~J., et al.\ 1995, \apj, 451, 188 

\bibitem[Silsbee et al.(2011)]{Silsbee:11} 
Silsbee, K., Ali-Ha{\"i}moud, Y., \& Hirata, C.~M.\ 2011, \mnras, 411, 2750 

\bibitem[Stepnik et al.(2003)]{Stepnik:03} 
Stepnik, B., Abergel, A., Bernard, J.-P., et al.\ 2003, \aap, 398, 551

\bibitem[Tibbs et al.(2010)]{Tibbs:10} 
Tibbs C.~T., Watson R.~A., Dickinson C. et al.,\ 2010, \mnras, 402, 1969 

\bibitem[Tibbs et al.(2011)]{Tibbs:11} 
Tibbs, C.~T., Flagey, N., Paladini, R., et al.\ 2011, \mnras, 418, 1889 

\bibitem[Tibbs et al.(2012a)]{Tibbs:12a} 
Tibbs, C.~T., Paladini, R., Compiegne, M., et al.\ 2012a, \apj, 754, 94 

\bibitem[Tibbs et al.(2012b)]{Tibbs:12b} 
Tibbs, C.~T., Scaife, A.~M.~M., Dickinson, C., et al.\ 2012b, submitted to \apj

\bibitem[Todorovi{\'c} et al.(2010)]{Todorovic:10} 
Todorovi{\'c}, M., Davies, R.~D., Dickinson, C., et al.\ 2010, \mnras, 406, 1629 

\bibitem[Vidal et al.(2011)]{Vidal:11} 
Vidal, M., Casassus, S., Dickinson, C., et al.\ 2011, \mnras, 414, 2424 

\bibitem[Watson et al.(2005)]{Watson:05} 
Watson R.~A., Rebolo R., Rubi{\~n}o-Mart{\'{\i}}n J.~A., Hildebrandt S., Guti{\'e}rrez C.~M., Fern{\'a}ndez-Cerezo S., Hoyland R.~J., Battistelli E.~S.,\ 2005, \apjl, 624, L89

\bibitem[Ysard \& Verstraete(2010)]{Ysard:10} 
Ysard, N., \& Verstraete, L.\ 2010, \aap, 509, A12 

\end{thebibliography}
\end{document}